\documentclass[twocolumn]{aastex7} %

\usepackage[utf8]{inputenc}

\usepackage{xspace}
\usepackage{amsmath}
\usepackage{amssymb}
\usepackage{mathptmx,txfonts,tikz,bm} 
\usepackage[multidot]{grffile}
\usepackage{xcolor, fontawesome}
\usepackage{graphicx}
\usepackage{physics}
\usepackage{comment}
\usepackage{color}
\usepackage{booktabs}
\usepackage{tabularx}
\usepackage{threeparttable}
\usepackage{multirow}
\usepackage{makecell}
\usepackage{alphabeta}

\newcommand{\mmode}[1]{\ifmmode{#1}\else{$#1$}\fi}

\newcommand{\Teff}[0]{\mmode{T_\text{eff}}}
\newcommand{\Rsun}[0]{\mmode{\text{R}_{\odot}}}
\newcommand{\Msun}[0]{\mmode{\text{M}_{\odot}}}

\newcommand{\Dnu}[0]{\mmode{\Delta\nu}}

\newcommand{\numax}[0]{\mmode{\nu_\text{max}}}

\newcommand{\fDnu}[0]{\mmode{f_{\Delta\nu}}}
\newcommand{\fnumax}[0]{\mmode{f_{\nu_{\rm max}}}}

\newcommand{\mh}[0]{\mmode{{\rm [M/H]}}}
\newcommand{\feh}[0]{\mmode{{\rm [Fe/H]}}}

\newcommand{\am}[0]{\mmode{{\rm [\alpha/M]}}}

\newcommand{\etar}[0]{\mmode{\eta_{\mathrm{R}}}}
\newcommand{\etasc}[0]{\mmode{\eta_{\mathrm{SC}}}}
\newcommand{\etam}[0]{\mmode{\eta_{\mathrm M}}}
\newcommand{\etacs}[0]{\mmode{\eta_{\mathrm{CS}}}}

\newcommand{\cyan}[1]{\textcolor{cyan} }

\usepackage{CJKutf8}

\makeatletter
\newcommand\thefontsize[1]{{#1 The current font size is: \f@size pt\par}}
\makeatother

\begin{document}
\begin{CJK}{UTF8}{gbsn}

\title{Evidence that Mass Loss on the Red Giant Branch Decreases with Metallicity}

\author[orcid=0000-0003-3020-4437]{Yaguang Li (李亚光)}
\affiliation{Institute for Astronomy, University of Hawai`i, 2680 Woodlawn Drive, Honolulu, HI 96822, USA}
\email[show]{yaguangl@hawaii.edu}

\begin{abstract}
Mass loss on the red giant branch (RGB) influences stellar evolution, properties of stellar populations, and Galactic chemical enrichment, yet remains poorly constrained observationally. Current models provide limited insight into how stellar properties, particularly how metallicity and mass, affect RGB mass loss. Here, I introduce a new observational approach that uses the age-velocity-dispersion relation and the lower-mass boundary of red giants as precise evolutionary markers. These markers, informed by Galactic evolution, allow us to construct observational isochrones for field stars. By comparing masses of RGB stars and red clump (RC) stars at the same age in the Kepler sample, I derive empirical measurements of integrated RGB mass loss at several points in age and metallicity. Combining these new observational measurements with open-cluster studies, I showed that the integrated mass loss on the RGB decreases with metallicity, and may also decrease with stellar mass. The average mass-loss rate, which accounts for RGB lifetimes and the initial mass differences between RGB and RC stars at the same age, also shows a similar trend. These findings challenge current mass-loss prescriptions widely adopted in stellar evolutionary models, since none of them is able to produce the observed mass-loss trend without widely adjusting free parameters. This highlights an urgent need to revise mechanisms that govern RGB mass loss. 
\end{abstract}


\section{Introduction}\label{sec:intro}

Mass loss during the red giant branch (RGB) phase is a critical but still poorly understood aspect of stellar evolution \citep{Catelan2009}. Low- and intermediate-mass stars ($M\leq2$~\Msun{}) typically lose up to 0.2~\Msun{} by the time they reach the RGB tip \citep{Pinsonneault2025}. This process has broad implications on inferred stellar ages for red clump (RC) stars \citep{Girardi2016}, horizontal branch morphologies in globular clusters \citep{Sweigart1997,Catelan2000}, and pulsation properties of RR Lyrae stars \citep{Caloi2008,Molnar2024,Zhanghy2025}. Mass loss also alters planetary system demographics and dynamics around evolved stars \citep{Veras2013,Villaver2014}. In close binaries, it shapes binary interactions at the RGB tip, affecting the formation of extreme horizontal-branch hot subdwarf-B stars \citep{Han2003}.

Integrated RGB mass loss has traditionally been inferred by comparing the mean mass of RGB and RC stars in the same stellar cluster, where they share the same age. This approach was most commonly applied in globular clusters through isochrone fitting to color-magnitude diagrams. It suggests a median mass loss of 0.2~\Msun{}, which also increase with metallicity \citep{McDonald2015,Tailo2021}. Independent mass measurements from asteroseismology confirmed these estimates \citep{Howell2022,Howell2024,Howell2025}. Similar conclusions arise from comparisons of field RGB and RC stars with $\alpha$-rich abundances. These stars originate from the thick disc, and are assumed to have similar ages \citep{Miglio2021,Brogaard2024}. In contrast, open clusters at near solar-metallicity such as NGC 6791, NGC 6819, and M67 show significantly lower mass loss, typically below 0.1~\Msun{} \citep{Miglio2012,Handberg2017,Stello2016,Reyes2025}, which are inconsistent with the metallicity trend observed in globular clusters.

Spectroscopic observations offer more direct diagnostics of RGB winds that drive mass loss. Features such as H$\alpha$, Ca II K, and He I lines often exhibit asymmetries or blue-shifted absorption, indicating chromospheric outflows \citep{Cacciari2004,Smith2004,Meszaros2009}. In luminous giants within globular clusters, wind velocities of tens to over 100~km/s have been detected, corresponding to mass-loss rates ranging from $10^{-9}$ to $10^{-7}$\Msun/yr that peak near the RGB tip \citep{Dupree2009}. Infrared observations can trace dust production, but enhanced dust emission is typically associated with asymptotic giant branch (AGB) stars rather than RGB stars, as seen in clusters such as 47~Tuc and $\omega$~Centauri \citep{McDonald2011,Boyer2010}. Dust emission also correlates with semi-regular pulsations, while RGB stars rarely show such features \citep{Yuj2021}.

Unlike the AGB mass loss --- where pulsation-driven, dust-enhanced winds are established to be responsible --- RGB mass-loss mechanisms remain elusive, thus the descriptions are largely empirical. Reimers’ law, for example, derives mass-loss rate by assuming a fixed fraction of stellar luminosity provides the lost gravitational potential \citep{Reimers1975}. While widely adopted, this relation does not directly describe the underlying physical processes responsible for the mass loss. Mass loss mechanisms generally include pulsation-driven winds, radiation pressure on dust and molecules, magnetic activity or Alfvén-wave driven winds, and binary-induced mass loss. Some prescriptions for the RGB incorporate Alfvén-wave physics \citep{Schroder2005,Cranmer2011,Suzuki2024}, but these have not been extensively tested in solar-metallicity or thin-disc populations due to a lack of identification for stars of same age. Studying a group of RGB and RC stars at the same age is necessary to characterize integrated RGB mass loss and, so far, such measurements have only been possible for the open clusters observed by Kepler.

In this study, I intend to address this critical gap by proposing new tools to identify field stars that are at the same age, which is informed by Galactic evolution. These tools include the age–velocity-dispersion (AVR) relation (\S\ref{sec:avr}) and the lower-mass boundary (LMB) of red giants (\S\ref{sec:lmb}). These methods offer extra constraints on RGB mass loss compared to current open cluster constraints (\S\ref{sec:oc}). When combined, they yield clear evidence that the integrated mass loss depends on metallicity and mass (\S\ref{sec:dmass}). In addition, I estimate the average mass-loss rates with stellar models (\S\ref{sec:rates}). Based on these constraints, I test common mass-loss prescriptions (\S\ref{sec:test}), discuss the implication for RGB mass-loss mechanisms and finally provide conclusions and outlook (\S\ref{sec:conc}).

\begin{figure}
    \centering
    \includegraphics[width=\columnwidth]{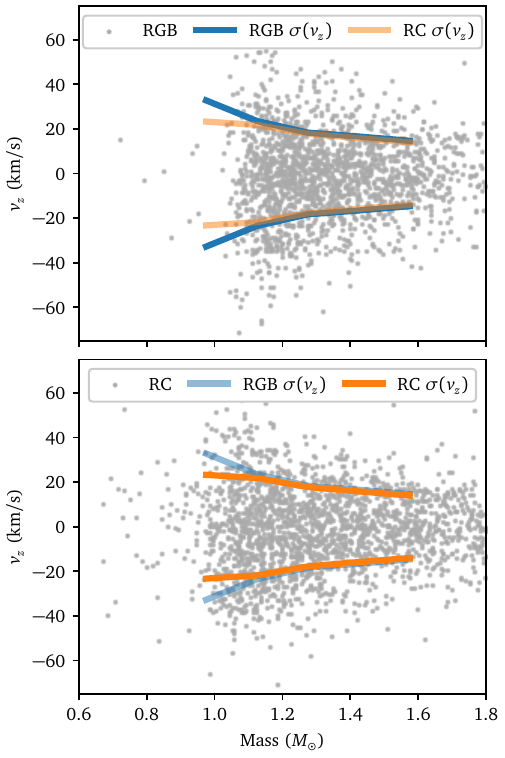}
    \caption{Vertical velocity $v_z$ vs. stellar mass for Kepler red giants in the thin disc characterized by $\alpha$-poor abundances. The top panel shows RGB and the bottom panel shows RC. The thick lines are standard deviations of $v_z$ in respective mass bins.}
    \label{fig:vz}
\end{figure}

\begin{figure}
    \centering
    \includegraphics[width=\columnwidth]{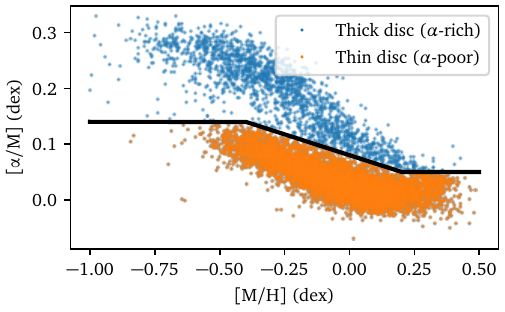}
    \caption{Classification of thick and thin disc stars based on the criterion adopted by \citet{Pinsonneault2025}.}
    \label{fig:class}
\end{figure}

\begin{figure*}
    \centering
    \includegraphics[width=\textwidth]{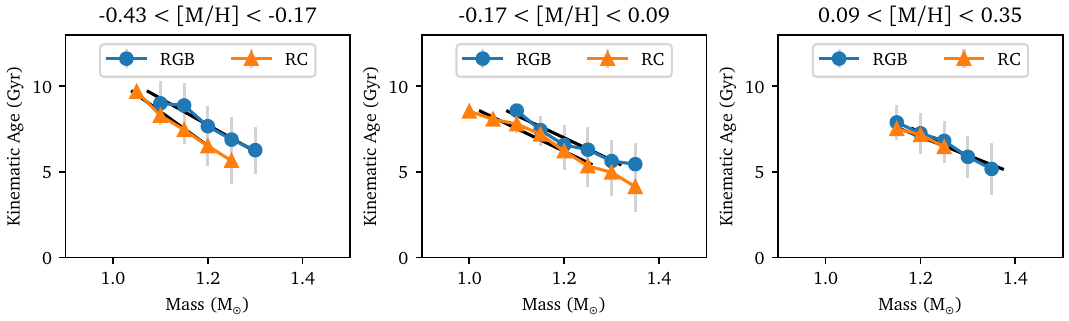}
    \caption{Kinematic age vs. stellar mass for Kepler red giants in the thin ($\alpha$-poor) disc. The black solid lines show the linear fit to the values for RGB and RC stars, respectively. }
    \label{fig:vd}
\end{figure*}

\section{Age-Velocity-dispersion relation}\label{sec:avr}

\subsection{Principle}\label{subsec:pp}

It has been established that older stars in the Galaxy generally exhibit larger velocity dispersions, while younger stars display smaller dispersions, as a result from dynamical heating over time. This kinematic velocity trend with stellar ages is known as the AVR (\citealt{Stromberg1946,Wielen1977}; see also \citealt{Bland-Hawthorn2016} for a review). 

Figure~\ref{fig:vz} illustrates this relationship by showing the vertical velocity $v_z$ (adopting a cylindrical coordinate system) as a function of stellar mass for RGB and RC stars observed by Kepler (see \S\ref{subsec:avr-data} for details on the sample selection). For illustration purposes, the standard deviations in separate mass bins are also shown. To first order --- without accounting for additional dependencies on metallicity or spatial position --- we can already observe that velocity dispersion increases with decreasing stellar mass \citep{Yujc2018,Anders2023}, which acts as a proxy for stellar age among red giants \citep{Wuyq2018,Schonhut-Stasik2024}. Additionally, at fixed mass, RC stars exhibit a smaller dispersion than RGB stars. This suggests that these RC stars are younger, consistent with their having higher initial masses, which were then reduced along the RGB.

In practice, we adopt the prescription proposed by \citet{Sharma2021}. In this framework, the vertical velocity $v_z$ follows a normal distribution with zero mean and standard deviation $\sigma_{v_z}$, expressed as
\begin{equation}\label{eq:vz}
    v_z \sim \mathcal{N}(0, \sigma_{v_z}).
\end{equation}
The dispersion $\sigma_{v_z}$ is expressed as a function of stellar age ($\tau$), vertical angular momentum ($L_z$), metallicity (\feh{}), and vertical coordinate ($z$): 
\begin{equation}\label{eq:sigmavz}
    \sigma_{v_z} = \sigma_{0,v_z} f_\tau f_{L_z} f_{\feh} f_z,
\end{equation}
where 
\begin{equation}\label{eq:ftau}
    f_\tau = \left(\frac{\tau/{\rm Gyr} + 0.1}{10 + 0.1} \right)^{\beta_{v_z}},
\end{equation}
\begin{equation}
    f_{L_z} = \frac{\alpha_{L,{v_z}}(L_z/L_{z,\odot})^2 + \exp \left[ -(L_z - L_{z,\odot}) /\lambda_{L,{v_z}} \right]}{1+\alpha_{L,{v_z}}},
\end{equation}
\begin{equation}
    f_{\feh} = 1+ \gamma_{\feh, {v_z}} \feh,
\end{equation}
and
\begin{equation}\label{eq:fz}
    f_z = 1+\gamma_{z, {v_z}} | z |.
\end{equation}
Here, the solar vertical angular momentum is fixed at $L_{z,\odot}=1935.36$~kpc~km~s$^{-1}$ \citep{Reid1993,Reid2004}.
\citet{Sharma2021} fitted this relation using field stars with well-determined ages, including main-sequence turn-off stars, and thereby obtained the following values of the free parameters: $\sigma_{0,{v_z}}=21.1\pm0.2$~km~s$^{-1}$, $\beta_{v_z}=0.441\pm0.007$, $\lambda_{L,{v_z}}=1130\pm40$~kpc~km~s$^{-1}$, $\alpha_{L,{v_z}}=0.58\pm0.04$, $\gamma_{\feh, {v_z}}=-0.52\pm0.01$~km~s$^{-1}$~dex$^{-1}$, and $\gamma_{z,{v_z}}=0.20\pm0.01$~km~s$^{-1}$~kpc$^{-1}$ (see Table 2 of \citealt{Sharma2021}).

Furthermore, by defining a variable $\mathcal{T}$ as 
\begin{equation}
    \mathcal{T}:= v_z / (\sigma_{0,v_z} f_{L_z} f_\feh f_z),
\end{equation}
we can obtain
\begin{equation}
    \mathcal{T} \sim {\mathcal N}(0, f_\tau),
\end{equation}
using Equations~\ref{eq:vz}--\ref{eq:sigmavz}. 
Assuming the mass loss process is either non-episodic or, if episodic, its effect can be averaged over evolutionary timescales, stellar age can be uniquely determined by mass, metallicity, and evolutionary stage. Based on this assumption, RGB and RC stars can be grouped into bins of metallicity and mass. In each bin, the standard deviation of $\mathcal{T}$ corresponds to $f_\tau$ and thus permits the estimation of kinematic stellar age using Equation~\ref{eq:ftau}.

\subsection{Data and Results}\label{subsec:avr-data}

The analysis in this paper is based on a sample of Kepler red giants from the APOKASC-3 catalog \citep{Pinsonneault2025}, which provides precise measurements of stellar masses, chemical abundances (including \mh{}, \am{}, and \feh{}), and evolutionary stages (RGB or RC). Stellar masses are derived from asteroseismic scaling relations. 

I restricted the sample to thin-disc stars because the thick disc spans a narrower age range, limiting the applicability of the AVR. Thin-disc stars were selected based on their \am{} and \mh{} abundances using the criterion adopted by \citet{Pinsonneault2025}. Figure~\ref{fig:class} shows the classification based on this criterion.

Spatial coordinates and velocities were calculated using the \texttt{galpy} package \citep{Bovy2015}, based on Gaia DR3 radial velocities, proper motions, right ascension, declination, and distances \citep{GaiaDR3-summary}.

Following the approach described in Section~\S\ref{subsec:pp}, I estimated kinematic ages for groups of stars defined by narrow bins in mass, metallicity, and evolutionary stage (RGB or RC). RGB and RC stars were grouped into overlapping bins of metallicity and mass to accommodate the limited sample size. Each mass bin has a width of 0.24~\Msun{} and a step size of 0.05~\Msun{}, while each metallicity bin has a width of 0.26~dex and a step size of 0.26~dex. As a result, the mass bins overlap, whereas the metallicity bins are independent.

Figure~\ref{fig:vd} shows the relationship between kinematic age and stellar mass. A clear correlation is observed, with RC stars exhibiting lower masses than RGB stars at a given age --- consistent with expectations of mass loss at the RGB tip. The mass difference,
\begin{equation}
    \Delta M = M_{\rm RGB} - M_{\rm RC},
\end{equation} 
provides an estimate of the integrated mass loss in a population of equal age. A clear trend emerges across metallicity bins: $\Delta M$ decreases with increasing metallicity. This metallicity dependence is discussed further in \S\ref{sec:dmass}. 

To determine $\Delta M$, I performed linear regressions to the age-mass relations shown in Figure~\ref{fig:vd} for RGB and RC stars separately. The regression was optimized using the following likelihood function:
\begin{equation}
    \ln p = \frac{1}{2}\sum_{i} (d_i - m_i )^2,
\end{equation}
where $d_i$ represents the observed ages, and $m_i$ denotes the modeled age derived from a linear function with mass, for the $i$-th data point.
The mass difference between the RGB and RC relations was calculated at a fixed age of $7.5$~Gyr. This age choice is not critical because the mass difference remains constant across ages due to the overlap of the mass bins. The uncertainties were estimated using a bootstrap method. Specifically, input measurements were randomly perturbed within their reported uncertainties, and the analysis was repeated over 500 trials. The standard deviation of the resulting $\Delta M$ estimates was adopted as the uncertainty. The final values are presented in Table~\ref{table:dmass}, in rows labeled AVR.

\begin{deluxetable}{lccccccc}
\tabletypesize{\footnotesize}
\tablecolumns{4}
\tablewidth{\textwidth} 
\tablecaption{Integrated mass loss on the RGB, calculated from the difference between the mean masses of RGB and RC stars in populations of equal age. \label{table:dmass}}
\tablehead{
\colhead{Methods} & 
\colhead{\hspace{0.1cm}$\Delta M$ (\Msun)}\hspace{0.1cm} & \colhead{\hspace{0.1cm}Mass (\Msun)}\hspace{0.1cm} & 
\colhead{\hspace{0.1cm}\mh{} (dex)}\hspace{0.1cm}  \\
}
\startdata
AVR & 0.06 $\pm$ 0.02 & 1.18 & -0.30\\
AVR & 0.06 $\pm$ 0.02 & 1.20 & -0.04\\
AVR & 0.00 $\pm$ 0.02 & 1.27 & 0.22\\
LMB $\alpha$-rich & 0.22 $\pm$ 0.02 & 0.99 & -0.70\\
LMB $\alpha$-rich & 0.14 $\pm$ 0.01 & 0.99 & -0.50\\
LMB $\alpha$-rich & 0.11 $\pm$ 0.01 & 1.04 & -0.30\\
LMB $\alpha$-rich & 0.08 $\pm$ 0.02 & 1.07 & -0.10\\
LMB $\alpha$-rich & 0.06 $\pm$ 0.02 & 1.11 & 0.10\\
LMB $\alpha$-poor & 0.05 $\pm$ 0.03 & 1.14 & -0.36\\
LMB $\alpha$-poor & 0.09 $\pm$ 0.02 & 1.22 & -0.18\\
LMB $\alpha$-poor & 0.06 $\pm$ 0.02 & 1.17 & 0.00\\
LMB $\alpha$-poor & 0.08 $\pm$ 0.02 & 1.24 & 0.18\\
LMB $\alpha$-poor & 0.01 $\pm$ 0.03 & 1.22 & 0.36\\
NGC 6791 & 0.02 $\pm$ 0.02 & 1.13 & 0.29\\
NGC 6819 & -0.02 $\pm$ 0.02 & 1.54 & 0.05\\
M67 & 0.00 $\pm$ 0.04 & 1.39 & 0.02\\
\enddata
\tablecomments{(1) AVR stands for age-velocity-dispersion relation and LMB stands for lower-mass boundary. (2) The values for M67 are adopted from \citet{Reyes2025}; other values are from this work. (3) Mass refer to the mass at the lower RGB.}
\end{deluxetable}

\begin{figure}
    \centering
    \includegraphics[width=\columnwidth]{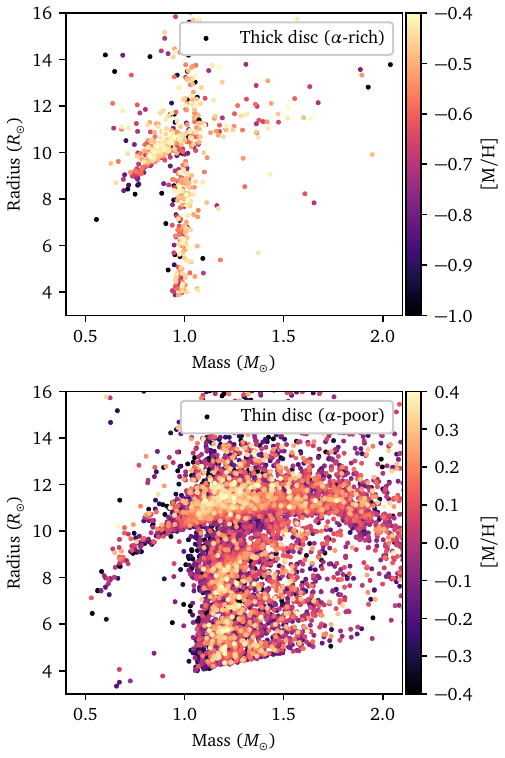}
    \caption{Mass-radius diagrams for Kepler red giants in both the thick ($\alpha$-rich) and thin ($\alpha$-poor) discs, color-coded by metallicity. }
    \label{fig:mr}
\end{figure}

\begin{figure}
    \centering
    \includegraphics[width=\columnwidth]{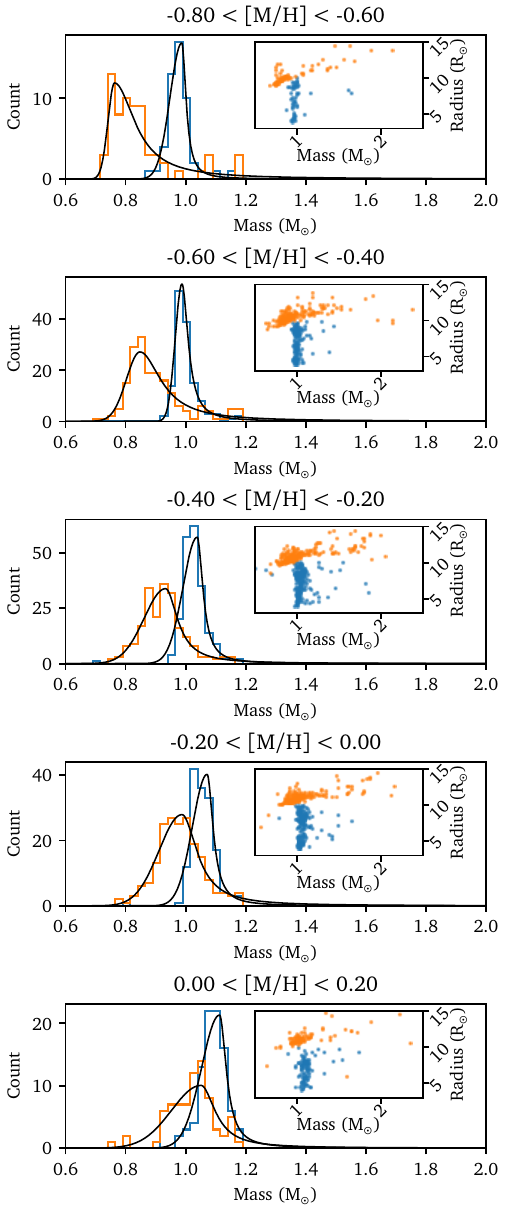}
    \caption{Mass distributions of the thick disc ($\alpha$-rich sequence) red giants, separated into metallicity bins. The distributions for RGB and RC stars are shown in different colors, each fitted using the parameterized function defined in Equation~\ref{eq:N}. The inset shows the corresponding mass-radius diagrams.
    }
    \label{fig:arich_edge_fit}
\end{figure}

\begin{figure}
    \centering
    \includegraphics[width=\columnwidth]{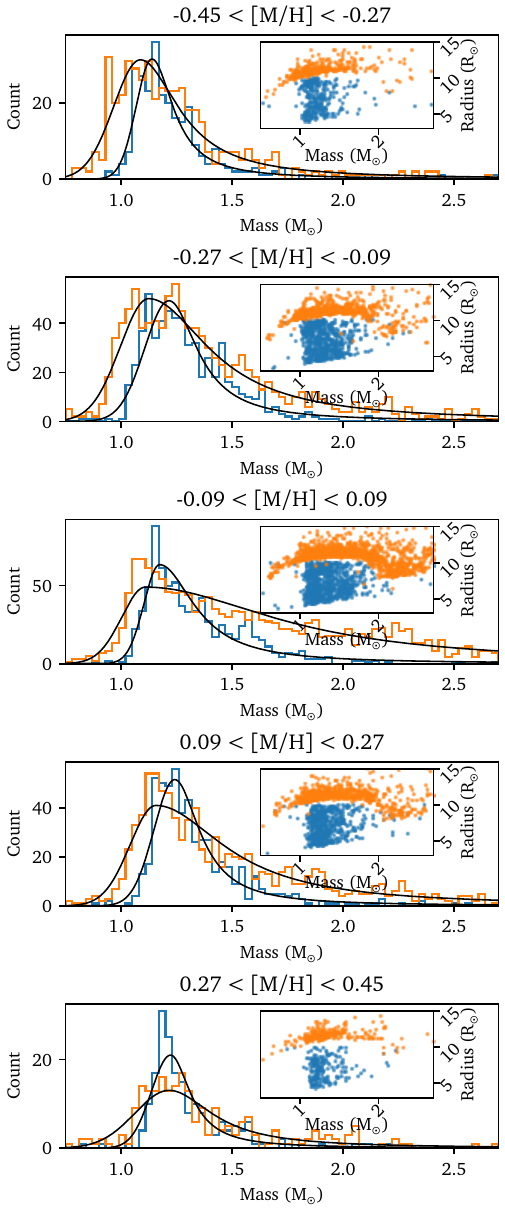}
    \caption{Mass distributions of the thin disc ($\alpha$-poor sequence) red giants, separated into metallicity bins. The distributions for RGB and RC stars are shown in different colors, each fitted using the parameterized function defined in Equation~\ref{eq:N}. The inset shows the corresponding mass-radius diagrams.}
    \label{fig:apoor_edge_fit}
\end{figure}

\section{Lower-mass Boundary}\label{sec:lmb}

\subsection{Principle}
Red giants span a broad mass range, but there is typically a lower-mass boundary, below which few are observed. This is because a star must have sufficient mass to evolve off the main sequence within the age of the Galaxy. For a given stellar population (e.g. a Galactic disc with a specified metallicity range), this lower-mass boundary effectively marks a fixed age, corresponding to the onset of the formation of that population.

\citet{Brogaard2024} applied a similar method to thick disc stars, using the mean mass rather than the lower-mass boundary. This can be considered as a valid approach given the approximate single-age (or, equivalently, single-mass) nature of the thick disc stars. For example, \citet{Miglio2021} reported a mean age of approximately 11 Gyr with a dispersion of about 1 Gyr for thick disc stars. See also \citet{Marasco2025} and \citet{Tayar2023} for a related application to metal-poor stars. They also estimated the integrated mass loss $\Delta M$ by comparing the mean masses of RGB and RC stars.

In more general cases --- particularly in the thin disc --- assuming a single-age population is not valid. Therefore, focusing on the lower-mass boundary, rather than on the mean mass, provides a more robust measure in such populations.

Figure~\ref{fig:mr} presents the mass-radius diagrams for Kepler red giants in the thick and thin discs. The curved horizontal feature comprises RC stars, which are restricted to a limited range of radii. The RGB stars cover a much wider range of radii. We see a distinct, sharp left boundary near 1~\Msun{}, particularly for the lower-RGB stars. Its position shifts with metallicity in a way that reflects the Galaxy's chemical enrichment history. The left boundary for RC stars is markedly lower than that for RGB stars, as a result of their having experienced mass loss. 

Additionally, in the thin disc the lower-mass boundary of RC stars is somewhat less sharp. It includes a group of sporadic, very low-mass outliers down to 0.5~\Msun{}. These are likely the result of enhanced mass loss due to binary interactions, as stellar envelopes are more likely to fill their Roche lobes near the tip of the RGB \citep{Liyg2022,Matteuzzi2023}. While these outliers are of separate interest, our analysis focuses on the main, dense population that presumably evolved through single-star processes.

\subsection{Method and Results}

I used the same Kepler red giant sample introduced in \S\ref{subsec:avr-data} and analyzed the thin and thick disc stars separately. Figures~\ref{fig:arich_edge_fit} and~\ref{fig:apoor_edge_fit} show the corresponding mass distributions for the RC and RGB populations, subdivided into metallicity bins. These distributions rise steeply with mass to a peak, followed by a more gradual decline. 

To model these distributions quantitatively, I fitted a parameterized function, $N(M; H, M_0, \sigma_M, \Gamma)$, to the stellar mass number counts. The left side of the distribution is described by a half-Gaussian profile, which approximates a convolved Dirac-$\delta$ function, while the right side follows a Lorentzian profile to capture the extended tail associated with the initial mass function. The adopted functional form is
\begin{equation}\label{eq:N}
    N(M) = \left\{ 
    \begin{aligned}
        & H \exp \left[-(M-M_0)^2/(2\sigma_M^2) \right], \ M \leq M_0; \\
        & \frac{H}{1 + (M-M_0)^2/\Gamma^2}, M > M_0.
    \end{aligned}
    \right.
\end{equation}
The parameter $M_0$ is interpreted as the lower-mass boundary. Poisson statistics were assumed for the number counts in each bin, resulting in the following likelihood function:
\begin{equation}
    \ln p = \sum_{m_i \neq 0} \left[ d_i \ln m_i - m_i - \ln (d_i !) \right],
\end{equation}
where $d_i$ and $m_i$ are the observed and model-predicted counts in the $i$-th bin. Uncertainties in the fitted parameters are derived from the sampling of this likelihood distribution.

Figures~\ref{fig:arich_edge_fit} and~\ref{fig:apoor_edge_fit} show the best-fitting functions based on Equation~\ref{eq:N}. The difference in $M_0$ between the RGB and RC populations provides an estimate of the integrated mass loss $\Delta M$, corresponding to the oldest stars within each metallicity bin. 
In some cases, the mass distributions exhibit slight deviations from a single Gaussian peak, likely due to random fluctuations caused by the small bin size. To evaluate its effect, I varied the bin size and found statistically consistent values for $\Delta M$.
These estimates are given in Table~\ref{table:dmass} in rows labeled LMB (lower-mass boundary).

\section{Open Clusters}\label{sec:oc}

Open clusters provide an ideal environment for studying stellar mass loss, because their members share a uniform age and initial chemical composition. Several open clusters observed by Kepler and K2 have yielded precise asteroseismic mass measurements, with NGC 6791, NGC 6819, and M67 being among the most extensively studied.

Using the APOKASC-3 catalog, I cross-matched stars in NGC 6791 and NGC 6819 based on the membership from \citet{Colman2022}. The integrated mass loss, $\Delta M$, was directly estimated from the difference between the mean masses of RC and RGB stars within each cluster. For NGC 6791, I derived an integrated mass loss of $\Delta M = 0.02\pm0.02~\Msun$ at an RGB mass of $M = 1.13~\Msun$. For comparison, \citet{Miglio2012} reported a broader range of $\Delta M$ values (0.02--0.19~\Msun), reflecting differences in the adopted scaling relations and reference mass scales. Their estimates were derived based on various combinations of luminosity, \Dnu{}, \numax{}, and \Teff{}, each introducing potential systematic uncertainties from the scaling relations or the reference parameters. The APOKASC-3 catalog reduces such inconsistencies by providing a uniformly calibrated dataset anchored to the Gaia radius scale. 

For NGC 6819, I found negligible mass loss, $\Delta M = -0.02\pm0.02 ~\Msun$, at $M = 1.54 ~\Msun$. This result is consistent with \citet{Handberg2017}, who reported $\Delta M = -0.03 \pm 0.01~\Msun$.

M67 was observed only with K2 and is not included in APOKASC-3. For this cluster, I adopted the measurements from \citet{Reyes2025}, who reported $\Delta M = 0.031 \pm 0.052~\Msun$ \citep[see also][]{Stello2016}, consistent with little or no measurable mass loss. The values for open clusters are also given in Table~\ref{table:dmass}.

\section{Integrated Red Giant Branch Mass Loss}\label{sec:dmass}

\begin{figure*}
    \centering
    \includegraphics[width=\textwidth]{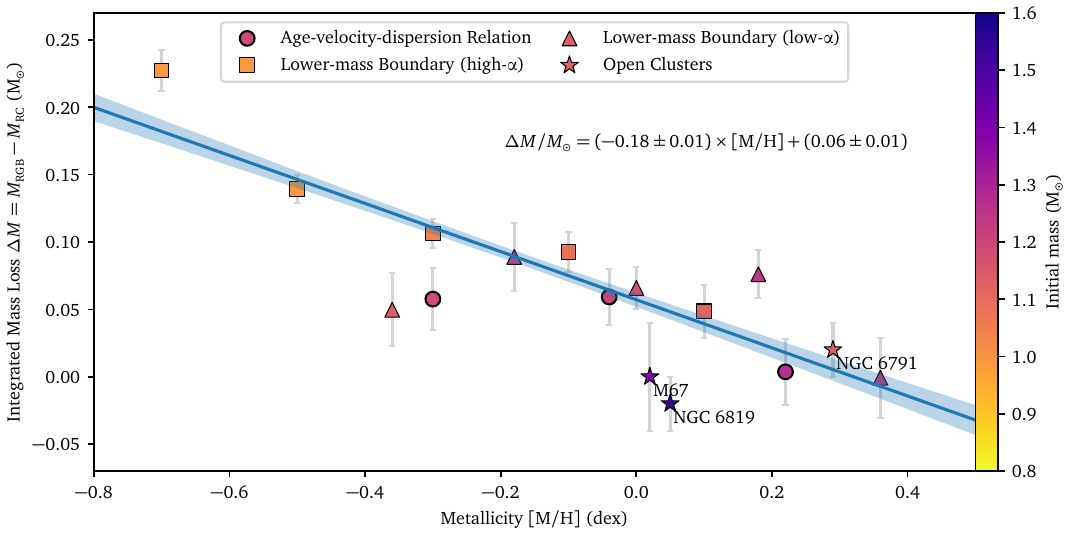}
    \includegraphics[width=\textwidth]{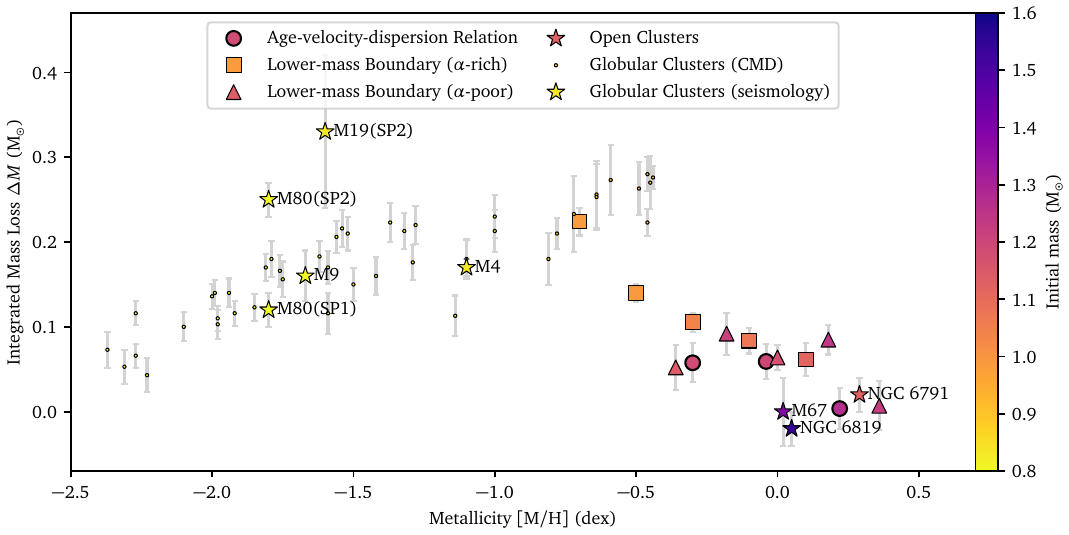}
    \caption{Top panel: integrated mass loss on the RGB obtained in this work, $\Delta M$, as a function of metallicity, color-coded by initial mass. A linear fit with a 1-$\sigma$ confidence region is shown. Bottom panel: same as the top panel, but including results from globular clusters \citep{Tailo2020,Howell2022,Howell2024,Howell2025}. }
    \label{fig:dmass}
\end{figure*}

\subsection{Trends with Metallicity and Mass}

The main results are presented in Table~\ref{table:dmass}, which lists the integrated mass loss, $\Delta M$, derived from the mass difference between RGB and RC stars at fixed age. Age estimates were based by the age–velocity relation (AVR), lower-mass boundary (LMB), or open cluster data discussed in \S\ref{sec:avr}–\ref{sec:oc}.

The top panel of Figure~\ref{fig:dmass} presents $\Delta M$ as a function of metallicity and initial stellar mass. A clear trend is evident: $\Delta M$ decreases with increasing metallicity. A linear fit to the data yields
\begin{equation}
\Delta M / \Msun{} = (-0.18 \pm 0.01)\times \mh + (0.06 \pm 0.01).
\end{equation}
There is also an indication that higher initial masses are associated with smaller $\Delta M$, a trend supported by the two open clusters M67 and NGC 6819. However, additional data are needed to confirm this mass dependence. This metallicity trend agrees with the findings of \citet{Brogaard2024}, who focused exclusively on $\alpha$-rich stars.

The metallicity-dependent mass-loss trend obtained from different methods or stellar populations generally shows good agreement. However, at \mh$\approx-0.4$, the values of $\Delta M$ estimated by the lower-mass boundary differ by $\approx2\sigma$ between $\alpha$-rich and $\alpha$-poor populations. This discrepancy may result from the less well-defined selection function of the Kepler sample \citep{Sharma2016}, which particularly affects $\alpha$-poor thin disc stars, making modeling with a Lorentzian tail less accurate. Future studies using samples with better-defined selection criteria, such as those from K2 and TESS, can help address this issue.

\subsection{Comparisons with Globular Clusters}
The field star and open cluster samples used in this work have metallicities above 0.8 dex. The results indicate that mass loss in the near-solar metallicity regime is modest, remaining below 10\% of the initial stellar mass.

How do these $\Delta M$ measurements compare to those for much older, more metal-poor globular clusters? The bottom panel of Figure~\ref{fig:dmass} presents $\Delta M$ measured for globular clusters by \citet{Tailo2020}, derived from the mass difference between the RGB tip and the horizontal branch (HB) using color-magnitude diagram morphology. It also includes $\Delta M$ values from \citet{Howell2022,Howell2024,Howell2025}, calculated by comparing masses between the lower RGB and either HB stars or early-asymptotic-giant-branch (eAGB) stars.

In these metal-poor globular clusters, RGB mass loss ranges from 0 to 0.3~\Msun{}. Its relationship with metallicity shows a positive correlation \citep{Gratton2010, Origlia2014, Tailo2020, Howell2025}, contrasting with the negative correlation observed in the near-solar metallicity sample \citep{Brogaard2024}.

However, these comparisons are not entirely straightforward. Globular clusters represent a much older population, and in some cases, the measured mass corresponds to AGB stars rather than stars at the onset of helium burning, which may lead to a slight overestimation of $\Delta M$. Furthermore, variations in individual chemical abundances beyond the bulk [M/H] likely affect mass loss, as different subpopulations within these clusters exhibit varying mass-loss amounts \citep{Tailo2020, Howell2025}. 

Despite these caveats, if the results are considered comparable, the non-monotonic trend of $\Delta M$ with [M/H] suggests that multiple mass-loss mechanisms may operate on the RGB. One may dominate in metal-poor environments, while another prevails at near-solar metallicities.

\begin{figure}
    \centering
    \includegraphics[width=\columnwidth]{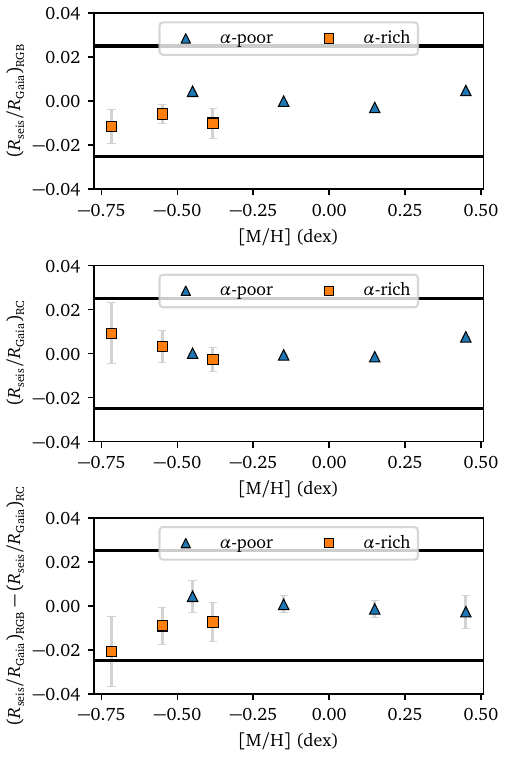}
    \caption{Differential comparisons between asteroseismic and Gaia radii for RGB and RC stars. The variations are smaller than 5\%, as indicated by the horizontal solid lines, which represents the level required to explain the $\Delta M$-\mh{} trend observed in Figure~\ref{fig:dmass}.}
    \label{fig:radius}
\end{figure}

\subsection{Accuracy of Asteroseismic Scaling Relations}

The $\Delta M$--\mh{} trend obtained in Figure~\ref{fig:dmass} depends critically on the accuracy of stellar mass estimates, which were derived from asteroseismic scaling relations by \citet{Pinsonneault2025}. Here, I investigate whether this trend could result from systematic biases in the mass scale rather than reflecting a physical effect.

The scaling relations rely on two key expressions: $\numax\propto\fnumax g T_{{\rm eff}}^{-1/2}$, and $\Dnu\propto\fDnu\sqrt{\rho}$,
where $g$ is surface gravity, $\rho$ is stellar mean density, and \fnumax{} and \fDnu{} are correction factors that are close to unity \citep{Ulrich1986,Brown1991,kb95,Sharma2016}. These expressions lead to mass and radius estimates through:
\begin{equation}
\label{eq:sc-mass}
    \frac{M_{\rm seis}}{\Msun} = \left(\frac{\numax}{\fnumax\numax_{,\odot}}\right)^3 \left(\frac{\Dnu}{\fDnu\Dnu_\odot} \right)^{-4} \left(\frac{\Teff}{\Teff_{,\odot}}\right)^{3/2},
\end{equation}
and
\begin{equation}
\label{eq:sc-radius}
    \frac{R_{\rm seis}}{\Rsun} = \left(\frac{\numax}{\fnumax\numax_{,\odot}}\right) \left(\frac{\Dnu}{\fDnu\Dnu_\odot} \right)^{-2} \left(\frac{\Teff}{\Teff_{,\odot}}\right)^{1/2}.
\end{equation}

A metallicity-dependent bias in Equation~\ref{eq:sc-mass} could, in principle, produce the observed trend in $\Delta M$. Two conditions must be satisfied for this to hold:
\begin{enumerate}
\item The scaling relation must yield different mass biases between the two evolutionary phases (RGB vs. RC).
\item The mass error, defined as $\delta (\Delta M) = \delta M_{\rm RGB} - \delta M_{\rm RC}$, must decrease by about 0.1~\Msun{} across the \mh{} range from $-0.4$ to $0.4$dex for $\alpha$-poor stars. This corresponds to a 10\% fractional error for a 1\Msun{} star. A similar 10\% decrease is needed for $\alpha$-rich stars from \mh{} $=-0.8$ to $-0.4$~dex.
\end{enumerate}

To evaluate the origins of possible errors, let us consider \numax{} and \Dnu{} separately. The quantity \numax{} is governed by surface processes related to mode excitation and damping operated in the stellar atmosphere. As a surface-driven quantity, there is limited justification for a dependence of the \numax{} relation on evolutionary phase.

The \Dnu{} scaling relation, however, depends on both evolutionary phase and metallicity, because it is tied to the overall structure of the acoustic cavity. Deviations from this relation are corrected using \fDnu{}, a theoretically derived factor, but it could be subject to uncertainties in stellar models. From Equations~\ref{eq:sc-mass} and \ref{eq:sc-radius}, fractional mass and radius errors relate to fractional \fDnu{} errors as $\delta M/M = 4 \delta\fDnu/\fDnu$ and $\delta R/R = 2 \delta \fDnu/\fDnu$, respectively, with each relation applying separately to RGB and RC stars. Therefore, a 10\% shift in $\delta (\Delta M)/M$ implies a 5\% shift in $\delta (\Delta R)/R$.

Because Gaia provides independent radius measurements, $\delta (\Delta R)/R$ can be estimated by $(R_{\rm seis} - R_{\rm Gaia})/R_{\rm Gaia}$, permitting a test of this hypothesis. I used $R_{\rm Gaia}$ values from the APOKASC-3 catalog, and computed the difference in this quantity between RGB and RC stars across metallicity bins, as shown in Figure~\ref{fig:radius}. The observed differences are at most 1\%, well below the 5\% required threshold, thereby ruling out this explanation. This conclusion assumes that the accuracy of the $\Dnu$ scaling relation is the sole factor contributing to the RGB-RC differences. I tested two additional mass and radius scales reported by \citet{Pinsonneault2025}, each using a different \fDnu{} correction. Both radius scales produced results consistent with those shown in Figure~\ref{fig:radius}.

Moreover, while asteroseismic scaling relations show signs of breakdown in very metal-poor regimes \citep{Chaplin2020,Huber2024,Larsen2025}, extensive testing near solar metallicity does not support such deviations \citep{Brogaard2018,Zinn2019,Liyg2021}.

\begin{figure*}
    \centering
    \includegraphics{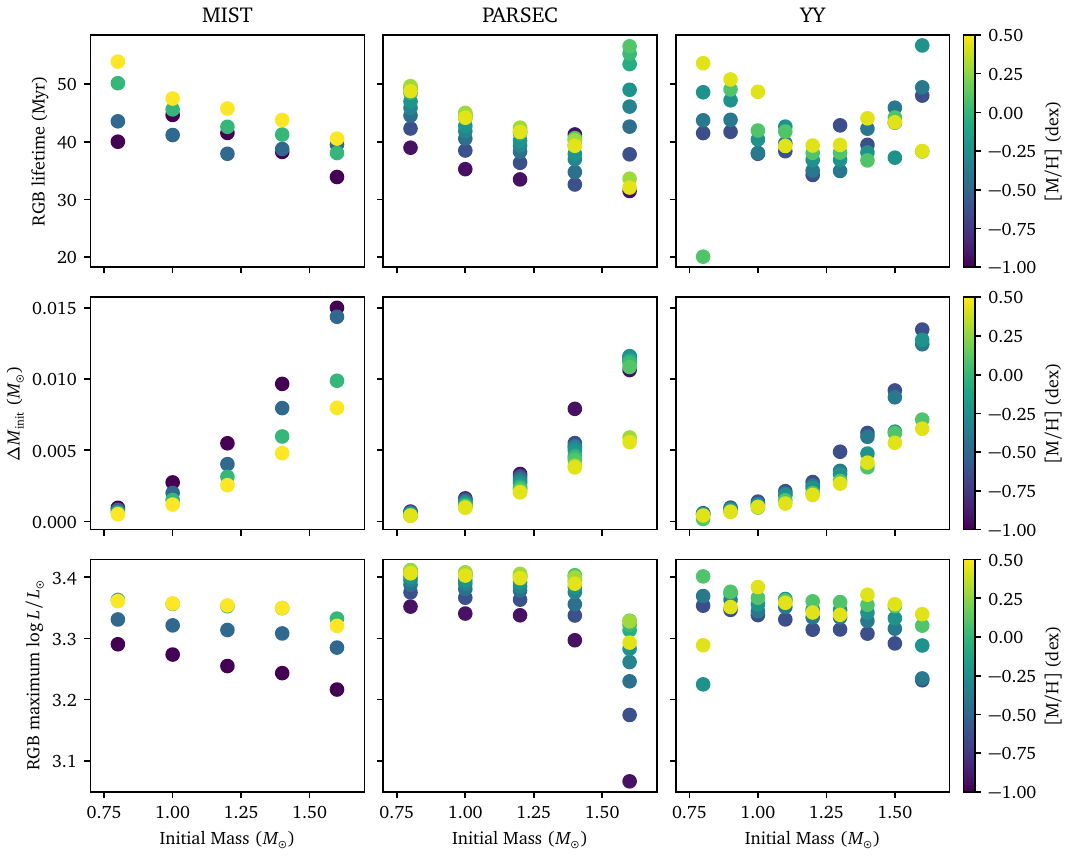}
    \caption{From top to bottom: RGB lifetime $\tau_{\rm RGB}$, initial mass difference $\Delta M_{\rm init}$ and maximum luminosity reached on the RGB, shown as functions of stellar mass and metallicity. Results are presented for three evolutionary models (left to right: MIST, PARSEC, and YY).}
    \label{fig:models}
\end{figure*}

\section{Average Mass-loss Rates}\label{sec:rates}
\subsection{Stellar Models}\label{subsec:models}

The integrated mass loss $\Delta M$ represents the total mass lost over the RGB phase, and depends on both the mass-loss rate and the RGB lifetime, $\tau_{\rm RGB}$.
Additionally, because RC stars are more evolved, their initial masses are slightly higher than those of RGB stars, which is a limitation when selecting samples of equal age. Correcting for this latter requires knowledge of the age–mass gradient, $d\tau/dM$. Further, if the mass-loss rate varies significantly along the RGB --- for example, if it increases steeply with luminosity --- then the maximum RGB luminosity, $L_{\rm RGB, max}$, should be used to consider possible implications.

I used stellar models to estimate the RGB lifetime ($\tau_{\rm RGB}$), the age–mass gradient ($d\tau/dM$), and the maximum RGB luminosity ($L_{\rm RGB, max}$). I used three commonly-used sets of stellar evolutionary tracks: MIST \citep[rotation version;][]{Choi2016}, PARSEC \citep[v1.2;][]{Chen2015}, and YONSEI-YALE \citep[YY;][]{Yi2001,Kim2002,Yi2003,Demarque2004}. These models cover metallicities from \mh{} = $-1.0$ to $0.5$ and masses from 0.8 to 1.6~\Msun{}. Using multiple models permits assessment of the robustness of any conclusions, given there are still uncertainties in RGB evolution.

First, I calculated the RGB lifetime, $\tau_{\rm RGB}$, as the time between the point where $R= 14$~\Rsun{} and the maximum luminosity reached on the RGB. As shown in the top panels of Figure~\ref{fig:models}, $\tau_{\rm RGB}$ depends primarily on metallicity for low-mass stars ($M \leq 1.6~\Msun$), decreasing from about 80~Myr at [M/H]~$\approx -1$ dex to roughly 40~Myr at [M/H]~$\approx 0.5$ dex. Its dependence on mass is comparatively modest, with variations below 10~Myr. All three models show these trends, though some scatter appears in the PARSEC and YY predictions, likely due to numerical issues. 

Next, I evaluated the initial mass difference between RC and RGB stars at fixed age, denoted $\Delta M_{\rm init}$. Because RC stars are more evolved, their progenitors had slightly higher initial masses. This difference can be approximated as
\begin{equation}\label{eq:dminit}
\Delta M_{\rm init} \approx \frac{ \tau_{\rm RGB}}{ d\tau/dM},
\end{equation}
where $d\tau/dM$ is the age–mass gradient. I fitted the age and mass of lower RGB models using the relation $\tau = \alpha M^\beta$, which gives $d\tau/dM = \alpha\beta M^{\beta - 1}$. Substituting this into Equation~\ref{eq:dminit} yields the estimate for $\Delta M_{\rm init}$. As shown in the middle panels of Figure~\ref{fig:models}, $\Delta M_{\rm init}$ increases with mass and decrease with metallicity, with a maximum of 0.025~\Msun{}. Overall, these values appear to be small compared to the integrated mass-loss rates determined in Table~\ref{table:dmass}.

I also determined the maximum RGB luminosity, as shown in the bottom panels of Figure~\ref{fig:models}. The overall variations are small, consistent with the fact that all stars in this regime undergo degenerate helium core formation followed by a helium flash. There is a  trend with metallicity, where metal-rich stars reach slightly higher RGB luminosities. However, if mass-loss rates indeed increase steeply with luminosity, this would imply that integrated mass loss should increase with metallicity, which is opposite to the trend observed.

\begin{figure*}
    \includegraphics[width=\textwidth]{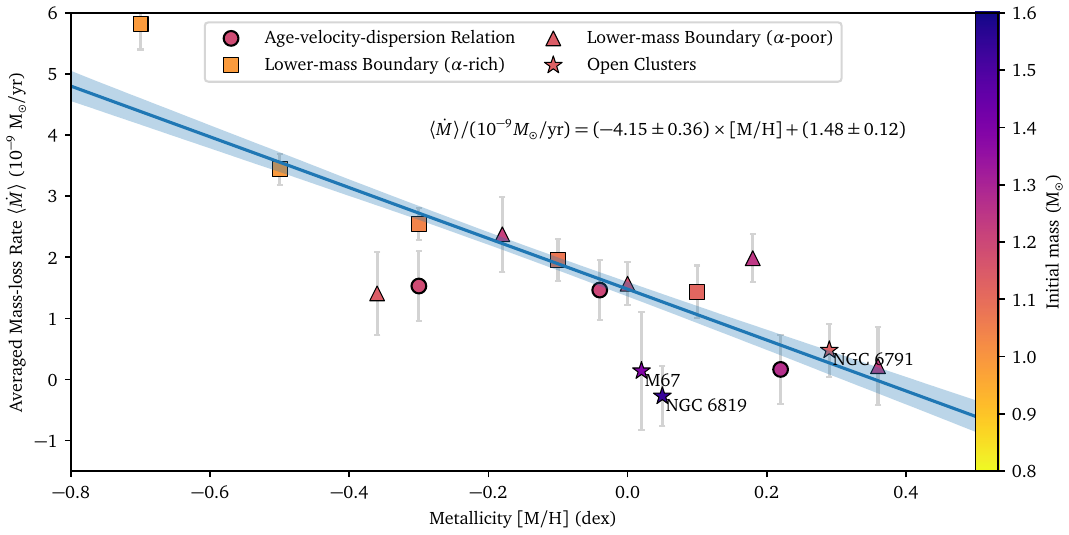}
    \caption{The average mass-loss rate on the RGB, $\langle \dot{M}\rangle$, as a function of metallicity, color-coded by initial mass. }
    \label{fig:dmass-rate}
\end{figure*}

\begin{figure*}
    \centering
    \includegraphics{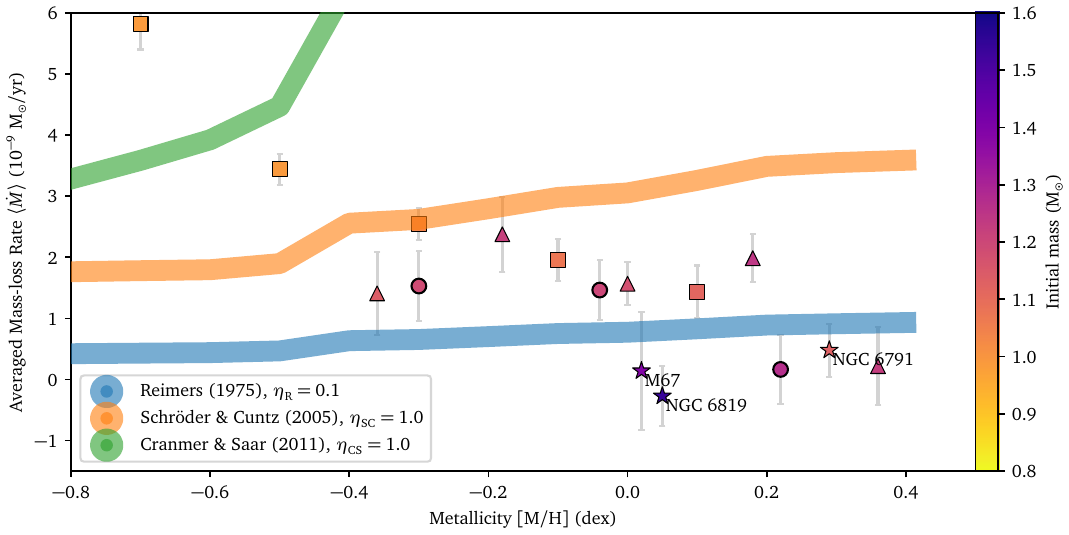}
    \caption{Predicted average mass loss rates from three different prescriptions. Observed values are shown as filled symbols, following the same conventions as in Figure~\ref{fig:dmass-rate}.}
    \label{fig:test}
\end{figure*}

\begin{figure}
    \centering
    \includegraphics{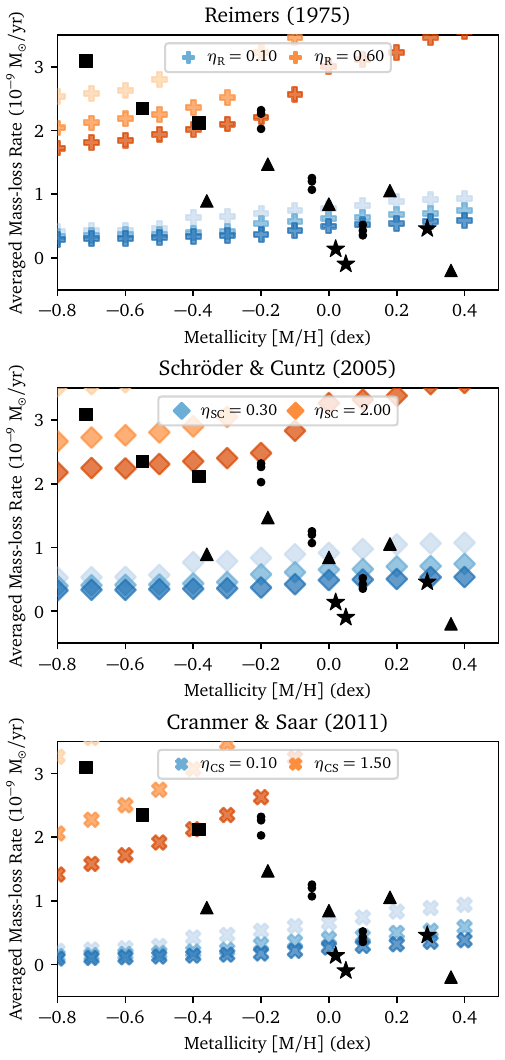}
    \caption{Predicted average mass-loss rates from three different prescriptions, shown as colored symbols. The color gradient represents stellar mass, from light to dark: 1.0, 1.2 and 1.4~\Msun{}. Observed values are indicated by black symbols, following the same conventions as in Figure~\ref{fig:dmass}. 
    }
    \label{fig:moretest}
\end{figure}

\subsection{Trends with Metallicity and Mass}
Combining the observationally derived integrated mass loss with the model-predicted RGB lifetime and initial mass differences allows us to estimate the average mass-loss rate on the RGB:
\begin{equation}
    \langle\dot{M}\rangle = \frac{\Delta M_{\rm true}}{ \tau_{\rm RGB}} = \frac{\Delta M + \Delta M_{\rm init}}{\tau_{\rm RGB}} = \frac{\Delta M}{ \tau_{\rm RGB}} + (d\tau/dM)^{-1}.
\end{equation}
For this calculation, I used the MIST models, since it exhibits smaller scatter and better numerical convergence. 

Figure~\ref{fig:dmass-rate} presents $\langle\dot{M}\rangle$ as a function of metallicity and stellar mass. The overall trends with metallicity remain consistent with those seen in the integrated mass-loss estimates (Figure~\ref{fig:dmass}). This is expected given that the values computed in \S\ref{subsec:models} have limited sensitivity to metallicity. Interestingly, the values of mass-loss rates agree qualitatively with those reported by \citet{Yuj2021}, who inferred mass-loss rates from infrared color excesses.
A linear fit to the data in Figure~\ref{fig:dmass-rate} yields
\begin{equation}
\begin{split}
\langle \dot{M} \rangle / & (10^{-9}M_\odot{\rm yr}^{-1}) \\
& = (-4.15 \pm 0.36)\times \mh + (1.48 \pm 0.12).
\end{split}
\end{equation}

\section{Testing the Mass-loss Prescriptions}\label{sec:test}

How do the measured mass-loss rates compare with model predictions? Here, I examine whether some commonly used mass-loss prescriptions reproduce the observed trends in the data, particularly the dependence of mass-loss rates on metallicity.

One of the most widely used mass-loss prescriptions is that of \citet{Reimers1975}. This model assumes that a fixed fraction (\etar{}) of the stellar luminosity balances the loss of gravitational energy due to stellar winds. Despite its broad application, the assumption of a constant fraction is a strong simplification, and the model does not incorporate detailed properties of stellar wind. The mass-loss rate is given by
\begin{equation}
\begin{split}
    \dot{M} = \ (4 \times 10^{-13} & M_\odot \  {\rm yr}^{-1}) \ \etar{} \\
& \left(\frac{L}{L_\odot}\right) 
\left(\frac{M}{M_\odot}\right)^{-1}
\left(\frac{R}{R_\odot}\right).
\end{split}
\end{equation}
Empirical calibrations suggest $\etar \approx 0.4$ for globular clusters and values in the range 0.1--0.2 for open clusters \citep{Miglio2012}.
Other prescriptions based on stellar fundamental parameters, such as those by \citet{Mullan1978}, \citet{Goldberg1979}, and \citet{Judge1991}, share a similar functional form and predict metallicity trends comparable to that of \citet{Reimers1975}. For this reason, we restrict our analysis to the \citet{Reimers1975} formulation.

Motivated by the mechanical energy flux carried by Alfvén waves, \citet{Schroder2005} proposed a modified mass-loss prescription:
\begin{equation}
\begin{split}
\dot{M} & = \ (8 \times 10^{-14} M_\odot \ {\rm yr}^{-1})  \ \eta_{\rm SC} \\
& 
\left(\frac{L}{L_\odot}\right) 
\left(\frac{M}{M_\odot}\right)^{-1} 
\left(\frac{R}{R_\odot}\right)
\left(\frac{\Teff}{4000~{\rm K}}\right)
\left(1 + \frac{g_\odot}{4300 g}\right),
\end{split}
\end{equation}
with $\etasc = 1$ originally chosen to reproduce the integrated mass loss observed in globular clusters \citet{Schroder2005}.

More recently, \citet{Cranmer2011} developed a mass-loss model based on solar wind physics. This model incorporates two distinct wind components driven by Alfvén waves and turbulence: a hot coronal wind \citep{Hansteen1995} and a cool wind originating from the extended chromosphere \citep{Holzer1983}. The total mass-loss rate is given by
\begin{equation}
    \dot{M} = \etam \left[ \dot{M}_{\rm hot} + \dot{M}_{\rm cold} \exp(-4M_{\rm A,TR}^2) \right],
\end{equation}
where $\dot{M}_{\rm hot}$ and $\dot{M}_{\rm cold}$ represent the contributions from the hot and cool wind components, respectively (see Section 3 of \citealt{Cranmer2011} for details). The term $M_{\rm A,TR}$ denotes the Mach number at the transition region between the hot corona and the cool chromosphere, and the scaling factor $\etam$ restores the original formulation when set to unity. Mass-loss rates were computed using the publicly available code\footnote{\url{https://stevencranmer.bitbucket.io/Data/Mdot2011/}}. Since the model requires the stellar rotation period as input, I estimated it using a radius-based scaling relation fitted from Table~2 of \citet{Cranmer2011}: $P_{\rm rot}/{\rm d} = 19.87\  (R/\Rsun)^{0.7}$.  

I computed the instantaneous mass-loss rates using the three prescriptions described above, applied to MIST stellar models. These rates were then integrated along the RGB and averaged over the RGB lifetime for direct comparisons with observational estimates. 

As a first step, I evaluated each prescription using its default scaling factor, restricting the analysis to models with an initial mass of 1.0~\Msun{} for simplicity. The results, shown in Figure~\ref{fig:test}, compare the predicted mass-loss rates from these prescriptions with the observed values found in this study. None of the models reproduce the observed metallicity dependence among field stars. The calibrated parameters in the \citet{Schroder2005} and \citet{Cranmer2011} prescriptions were primarily derived from metal-poor samples, and the poor agreement near solar metallicity is unsurprising.

Next, I explored the effect of varying the scaling factors --- \etar{}, \etasc{}, and \etacs{} --- within each prescription. In this case, I also included models with three different initial masses: 1.0, 1.2, and 1.4~\Msun{}. As shown in Figure~\ref{fig:moretest}, this additional freedom does not resolve the discrepancy. Although all prescriptions produce a mass gradient with a sign consistent with the observed trend with mass \citep{TheanoTheodoridis2025}, the overall agreement remains poor.

\section{Discussions and Conclusions}\label{sec:conc}

In this study, I present constraints on the integrated mass loss on the RGB using Kepler red giants. This was done by comparing stellar masses of RGB and RC stars of equal age, which were identified using two Galactic evolution indicators: age–velocity dispersion relation (AVR) and lower-mass boundary (LMB) of red giants. Together with measurements from open clusters, I showed that the integrated mass loss on the RGB clearly decreases with metallicity for $\mh > -0.8$~dex, and also tentatively decreases with stellar mass for $M<1.6~\Msun{}$ (Figure~\ref{fig:dmass}).

The findings presented here have broad implications across astrophysics. They indicate that current stellar evolutionary models systematically misrepresent mass-loss behavior in evolved stars, affecting age estimates for stellar populations --- particularly for red clump stars, which are important for galactic archaeology. Interpretations of red clump morphology and models of Galactic chemical evolution must account for the metallicity dependence of mass loss when adopting empirical prescriptions.

The trend between mass loss and metallicity starkly contrasts with existing widely adopted mass-loss prescriptions, including \citet{Reimers1975}, \citet{Schroder2005}, and \citet{Cranmer2011}, since none of them is able to reproduce the observed metallicity dependence, even when allowing generous adjustments to their free parameters (Figures~\ref{fig:test} and~\ref{fig:moretest}).

The results presented here also have implications for the physical mechanisms responsible for mass loss on the RGB. Pulsation-enhanced mass loss have been widely recognized as a mechanism in evolved stars, particularly in luminous AGB stars, where pulsations can levitate material to regions where dust can condense. Yet, while the tip of the RGB may just reach pulsation amplitudes similar to those seen in AGB stars, the empirical finding that pulsation amplitudes increase with metallicity in RGB stars \citep[see Figure~12 of][]{Yuj2018} would suggest, at face value, that mass-loss rates should also increase with metallicity --- opposite to the trend observed in this work.

Radiation pressure on dust grains is a well-established driver of winds in hot main-sequence stars and very luminous AGB stars. However, in RGB stars the lower luminosities and cooler atmospheres significantly limit both the formation and effectiveness of dust-driven winds.

Magnetic fields present a promising, though less well-understood, mechanism for driving mass loss. Although the prescriptions by \citet{Schroder2005} and \citet{Cranmer2011} are based on Alfvén wind physics, their implementations rely on simplifying assumptions. Magnetohydrodynamic (MHD) simulations suggest that a range of wind regimes may arise from magnetic activity, but the dependence on metallicity remains poorly constrained \citep{Yasuda2019}. The slowly and, potentially, differentially rotating envelopes of red giants complicate direct comparisons with dynamo processes observed in solar-like stars. Future measurements of surface rotation periods in red giants, along with envelope and core rotation rates from asteroseismology \citep{Lig2024,Hatt2024}, should help clarify the role of differential rotation. In parallel, magnetic activity indicators along the red giant branch are essential for constraining surface magnetic field strengths, with recent progress in this area \citep{Gehan2024}. Together, these efforts are expected to improve our understanding of the interaction between rotation and magnetic fields in red giants, and may help clarify the role of magnetic activity in mass loss.

On the observational side, it is worth noting that current isochrone sequences informed by Galactic evolution indicators primarily favor low-mass stars. The only intermediate-mass samples are provided by the open clusters M67 and NGC 6819. To robustly assess the role of stellar mass by extending the parameter space toward higher masses, additional open cluster observations are needed --- although the number of suitable clusters in the field is limited. A promising alternative involves the use of wide binary systems, where the primary is an RC star and the companion is either an RGB or a main-sequence turn-off star with a well-determined age. Preliminary results from such studies have shown considerable potential \citep{Schimak2024,Chiu2025}.

In addition, stellar masses derived from asteroseismic scaling relations are subject to systematic uncertainties in the absolute mass scale. Individual frequency modeling provides another promising, though currently underutilized, alternative for obtaining precise mass estimates. Applying detailed frequency modeling to large samples of red clump stars should be prioritized to improve constraints on mass-loss processes.

\section*{Acknowledgements}
I am grateful to Daniel Huber and Tim Bedding for reading and commenting on the manuscript, to Neige Frankel and Jennifer van Saders for discussions at earlier stage of the project, and to the referee for their time and helpful comments. This work is supported by the Beatrice Watson Parrent Fellowship.

\end{CJK}

\bibliography{references/myastrobib}{}
\bibliographystyle{aasjournal-compact}

\end{document}